# Viewing short Gamma-ray Bursts from a different angle

J. Michael Burgess[1,2], Jochen Greiner[1], Damien Begue[1], Dimitrios Giannios[3], Francesco Berlato[1], Vladimir M. Lipunov[4]

[1] Max-Planck Institut für Extraterrestrische Physik, Giessenbachstr. 1, 85748 Garching, Germany
[2] Excellence Cluster Universe, Technische Universität München, Boltzmannstr. 2, 85748, Garching, Germany
[3] Department of Physics and Astronomy, Purdue University, 525 Northwestern Avenue, West Lafayette, IN 47907, USA
[4] Sternberg Astronomical Institute, Lomonosov University, 119899 Moscow, Russia

**The recent coincident detection of gravitational waves (GW) from a binary neutron star merger with aLIGO/Virgo[1,2,3] and short-lived gamma-ray emission with Fermi/GBM[4,5] (called GW 170817) is a milestone for the establishment of multi-messenger astronomy. Merging neutron stars (NS) represent the standard scenario for short-duration (< 2 sec) gamma-ray bursts[6] (GRBs) which are produced in a collimated, relativistically expanding jet with an opening angle of a few degrees and a bulk Lorentz factor of 300-1000. While the present aLIGO detection is consistent with predictions, the measured faint γ-ray emission from GW 170817A, if associated to the merger event at a distance of 40 Mpc, is about 1000x less luminous than known short-duration GRBs (sGRBs)[7]. Hence, the presence of this sGRB in the local Universe is either a very rare event, or points to a dramatic ignorance of the emission properties of sGRBs outside their narrow jets. Here we show that the majority of previously detected faint sGRBs are local, at redshift smaller than 0.1, seen off-axis. In contrast, the brighter sGRBs are seen on-axis, and therefore out to larger distances, consistent with the measured redshift distribution[7]. Examining the observer-frame parameter space of all Fermi/GBM sGRBs shows that the sGRB associated with GW 170817A is extreme in its combination of flux, spectral softness and temporal structure. We identify a group of similar GRBs, one of which has been associated to a bright galaxy at 75 Mpc. We incorporate off-axis emission in the estimate of the rates of sGRBs, and predict that the majority of future GW-detections of NS-NS mergers will be accompanied by faint γ-ray emission, contrary to previous thinking. The much more frequent off-axis emission of sGRBs also implies a much higher deadly rate of γ-rays for extraterrestrial life in the Universe.**

Fermi/GBM triggered on a burst-like transient on August 17, 2017 at 12:41:06.5 UT (trigger 524666471 / 170817529)[4,5]. With a GBM localization[4,5] consistent with that of the aLIGO trigger[2] about 2 seconds earlier, there is little doubt that these two triggers are from the same event, a neutron star binary coalescence[1,3].

The burst consists of a single pulse. The event fluence during the ~0.65 sec prime emission time interval in the 10-1000 keV energy range is $(2.45\pm0.4)\times10^{-7}$ erg cm$^{-2}$. Combined with the LIGO distance estimate of 40 Mpc[1,8] this implies an isotropic equivalent energy of $5.23\times10^{46}$ erg in the 1 keV – 10 MeV band (this energy band is given for comparability; no photons above 400 keV are measured for this event). The count data of the single significant bin from the event was fit to an exponentially cutoff power law resulting in a photon spectral index of $\alpha = -0.98\pm0.2$ and cutoff energy of $E_c = 300_{-150}^{+290}$ keV. Comparing the spectral properties of GW 170817 to the sample of previously detected GRBs[9] by Fermi/GBM shows that it is not a typical short GRB. The event resides on the boundary of the $\alpha$-flux, $\alpha$-$E_c$ plane and temporal smoothness, but is accompanied by a few sGRBs with similar properties (Fig. 1).

While there is only one published prediction of the properties of off-axis emission[10], the generic expectations are a decreased luminosity and non-complex observed light curve structure due to the relativistic smoothing of the internal emission episodes[11]. Similarly, the stacking of spectra observed over the profile of the jet leads to an observed increase in low-energy photons and thus a softer spectrum below the $\nu F \nu$ peak[12]. Conversely, observed on-axis emission probes the internal Lorentz profile for the jet allowing the observer to see multi-episodic emission as well as the direct spectral shape of the spectrum.

Accepting the conjecture that for GW 170817 we detected gamma-ray emission far off-axis with the above properties, we searched our GBM GRB database for events with similar spectral and temporal characteristics. On December 24, 2010, at 05:26:57.5 UTC, both Swift/BAT and Fermi/GBM triggered on the short GRB 101224A, lasting 0.36 s. Our analysis of its temporal and spectral properties shows that the event is remarkably similar to GW 170817 including its light curve structure. As noted earlier[13], there is the galaxy pair MCG+08-34-033 nearby, at a distance of 75 Mpc. The offset of this GRB to the center of MCG+08-34-033 is <135 kpc, not unusual for sGRBs[14]. Assuming this to be the distance of the sGRB, we calculate a fluence of $2.8\times10^{47}$ erg in the 1 keV – 10 MeV band. For comparison, the short GRB 09015A at distance of 5990 Mpc has a fluence of $2.7\times10^{53}$ erg and GRB 100625A at a distance of 2590 Mpc has a fluence of $2.3\times10^{53}$ erg, making them markedly dissimilar from the two local events we identify. In Fig. 1, the two candidate off-axis GRBs are spectrally weak and soft in comparison to the population, however, they do not deviate from the typical cutoff energy distribution. In contrast, GRB 0090510A and GRB 100625A are spectrally harder and brighter while being two orders of magnitude further away.

With another promising off-axis sGRB identified, we can ask the question: how many of these events we have detected with GBM in the past without recognizing them as being off-axis? In order to estimate the rate of sGRBs, we follow simple estimates[15,16]. After determining a maximum distance up to which the GRB is assumed to be detected, the rate of GRB 170817A and GRB 101224A can be estimated to be ~170 Gpc$^{-3}$ yr$^{-1}$ and 2 Gpc$^{-3}$ yr$^{-1}$. This is far more than the rate of individual other short GRBs, being around 0.03 Gpc$^{-3}$ yr$^{-1}$, with the exception of GRB 080905. Note that the rates strongly depend on the chosen flux limit chosen to describe the sensitivity of GBM. Yet, a direct comparison with GRB 100625A and GRB 09051A, show that their rates are very high.

In Fig. 2, on and off-axis emission rates are separated to show where each component of the total rate would dominate. For the local Universe, off-axis emission would dominate. Notably, after integrating deep into the Universe, the low end of the observed flux distribution would be dominated by off-axis events. Therefore, one can wonder if such GRBs are part of a new undiscovered population of sGRBs, comparable to the population of under-luminous long GRBs, or if they are members of the same population, but seen off-axis. Indeed, one naturally expects the rate of local short weak bursts to be dominated by off-axis ones. The ratio of such off-axis bursts to on-axis bursts strongly depends on the structure of the jet and on the luminosity function; both poorly understood. In the supplementary material, we estimate the ratio of sGRBs seen on- and off-axis as a function of redshift. We find that the rate of off-axis sGRBs at redshift smaller than 0.02 make up the bulk of the observed distribution. Conversely, on-axis sGRBs dominate the observed rate at high redshift. Even if this computation needs to be refined, it indicates that most of the sGRBs seen by aLIGO should be observed off-axis. Thus, on-axis sGRBs beamed at us are visible to a redshift of about 1.5 (the highest redshift sGRBs are 1.4, 1.7 and 2.6, the last one being highly debated); off-axis GRBs are unbeamed, and with the same instrument sensitivity can only be seen to about 1/30 of those distances, i.e. about 300 Mpc . Our analysis of GBM data suggests that at least a few percent, if not up to 30%, of sGRBs are local. This is fully consistent with the detection of one such event during aLIGO run O2 with a distance threshold of 100 Mpc. Given the assumption that the bulk of sGRBs are local off-axis events, it begs the question to why this is not observed in the Swift sGRB redshift distribution? It is interesting to hypothesize that a large observational bias has occurred. Faint X-ray afterglows are observed by Swift/XRT, often not proven to fade, assumed to be associated with a host galaxy, and then not followed up. If many of these afterglows are in fact persistent sources, unassociated with the sGRB event, then the observed Swift sGRB redshift distribution could have contributed in a beautiful way to fool astronomers.

How could off-axis emission be generated? Relativistic motion is fundamental to GRBs. The first light emitted by the relativistically expanding plasma originates from the photosphere, the surface from which photons decouple from the outflow. The delay between the aLIGO event and the associated sGRB can be used to set a lower limit on the Lorentz factor $\Gamma$ of the outflow expanding towards the observer by considering the time it takes for the outflow to reach the photosphere. We find $\Gamma > 3$. Note that the limit on the Lorentz factor is higher if the result of the merger is an unstable neutron star, collapsing shortly after the GW event to a black-hole, or if the gamma-ray emission takes place beyond the photosphere of the jet, through shocks or magnetic reconnection.

The merger is accompanied by ejection of neutron-star material in the surrounding space[17]. The importance of this ejecta is twofold. On the one hand, it helps to collimate the relativistic jets by ram pressure[18]. On the other hand, when the jet breaks out the slower ejecta, a substantial amount of its material is heated and accelerated, forming a cocoon[19,20]. The expansion of this cocoon is nearly isotropic, as it experiences lateral expansion, up to an angle of ~50°. The cocoon might power a short gamma-ray transient when it becomes transparent. Assuming that the emission of the cocoon is thermal, estimations show that the peak energy of the spectrum is ~1 keV[19], pointing towards the need of additional non-thermal processes to explain the break energy around 100 keV for GW 170817. Magnetic dissipation at the slower, high-latitude, jet can provide the required particle acceleration[21].

While the progenitors of sGRBs have long been speculated to be the merger of compact remnants, the reservoir of energy powering their gamma-ray emission is still debated. The delay between the GW signal and the gamma-ray emission allows one to speculate on the product of the merger, and as a result, the origin of the energy. If the binary is massive enough, the merger results instantly in the creation of a black-hole, with accretion of any remaining material powering the gamma-ray burst. Accretion could also power the sGRB if the result of the merger is an unstable neutron star, collapsing to a black-hole within the 2 seconds delay between the sGRB and the GW event. However, if the result of the merger is a (quasi-) stable neutron star, the gamma-ray burst emission would be delayed. Indeed, strong neutrino losses from the cooling newly formed neutron star may drive strong mass losses, preventing relativistic jets to be launched until 5 to 10 seconds after the merger[22].

The combined detection of gravitational waves and relatively faint gamma-ray emission from a binary neutron star merger implies that either the previously accepted distance scale to sGRBs is wrong, or that we were not within the beam of GW 170817. With 35 out of 104 short GRBs measured with Swift having redshifts and all of these redshifts being larger than 0.1, it looks like a contrived assumption to assume all other sGRBs are below z<0.1, down to a few dozen Mpc. We have thus investigated the off-axis option. With realistic assumptions on the opening angle distribution of the jetted emission, and a low-level off-axis emission we predict that at fluxes below $10^{-5}$ erg cm$^{-2}$ s$^{-1}$ the observed rate of sGRBs is dominated by local NS-NS binaries seen off-axis. This prediction is confirmed by the Fermi/GBM data which reveal a group of faint sGRBs with similar spectral and timing properties (Fig. 1), suggesting that the above alternative explanation is not that contrived at all. While the observed single NS-NS merger at 40 Mpc within the aLIGO O2 run is grossly consistent with previous predictions of the merger rate in the local surrounding[23,24,25], we conclude from Fig. 2 that during O3 we can expect many common aLIGO/GBM detections.

Past campaigns to follow-up sGRBs have focused their efforts on bright, energetic events. If we liberally extend our conclusions that dim, and spectrally uninteresting events are in fact signatures of local neutron star mergers, then we must refocus our follow-up campaigns to examine and learn more about them. With a sizable sample of confirmed local events, better estimates on rates will follow, and thus more accurate predictions for LIGO/VIRGO detections. Additionally, it has already been estimated that GRBs, in general, leave only ~10% of galaxies hospitable for life and only after z<0.5[26,27]. A larger local population of sGRBs would substantially lower this rate and bring it closer to z=0. Therefore, observations to confirm the existence of local sGRBs will aid in answering the question of are we really alone in the Universe?

**Supplementary Information** is linked to the online version of that paper at www.nature.com/nature.

**Acknowledgements.**
JMB and JG acknowledge support by the DFG cluster of excellence "Origin and Structure of the Universe" (www.universe-cluster.de). DB, FB and JG acknowledge support by the DFG through SFB 1258. DG acknowledges support from NASA through the grants NNX16AB32G and NNX17AG21G issued through the Astrophysics Theory Program. AvK is supported by the Bundesministerium für Wirtschaft und Technologie (BMWi) through grant 50 OG 1101 from the Deutsches Zentrum für Luft- und Raumfahrt (DLR). Support for the German contribution to Fermi-GBM was provided by the Bundesministerium für Bildung und Forschung (BMBF) via DLR under contract number 50 QV 0301. The MASTER project is supported in part by the Development Program of Lomonosov Moscow State University, Moscow Union OPTICA, Russian Science foundation grant 16-12-00085.

**Author contributions.** JMB has performed data analysis and catalog searches for candidate gamma-ray events. JG suggested the difference to the sGRB population, and together with VML have performed archival data searches for counterparts to sGRBs. DB, DG and VML have been instrumental in various aspects of the data interpretation. All have been participating in the write-up and discussion of the conclusions.

**Author Information.** Reprints and permissions information is available at www.nature.com/reprints. The authors declare no competing financial interests. Correspondence and requests for materials should be addressed to jburgess@mpe.mpg.de.

**Figure 1: Observed spectral properties of GW 170817.**
Comparison of the spectral properties of GW 170817 and the spectrally and temporally similar GRB 101224A with respect to our sample of well-fit short GRBs. Colors indicate light curve structure (see labels in the top left) and size indicates the duration from our Bayesian block analysis. The shaded region corresponds to the flux level at which off-axis short GRBs become the dominant contribution to the total rate (see Figure 2).

**Figure 2: Short-GRB rate distribution with respect to observed flux.**
The short-GRB rate distribution as a function of flux calculated by assuming a jet structure and luminosity function and then consoling this with the binary neutron star merger rate for various redshift integration ranges. The GRBs with redshift in our sample are plotted for comparison. The shaded region indicates the flux values for which GBM has not observed a GRB.

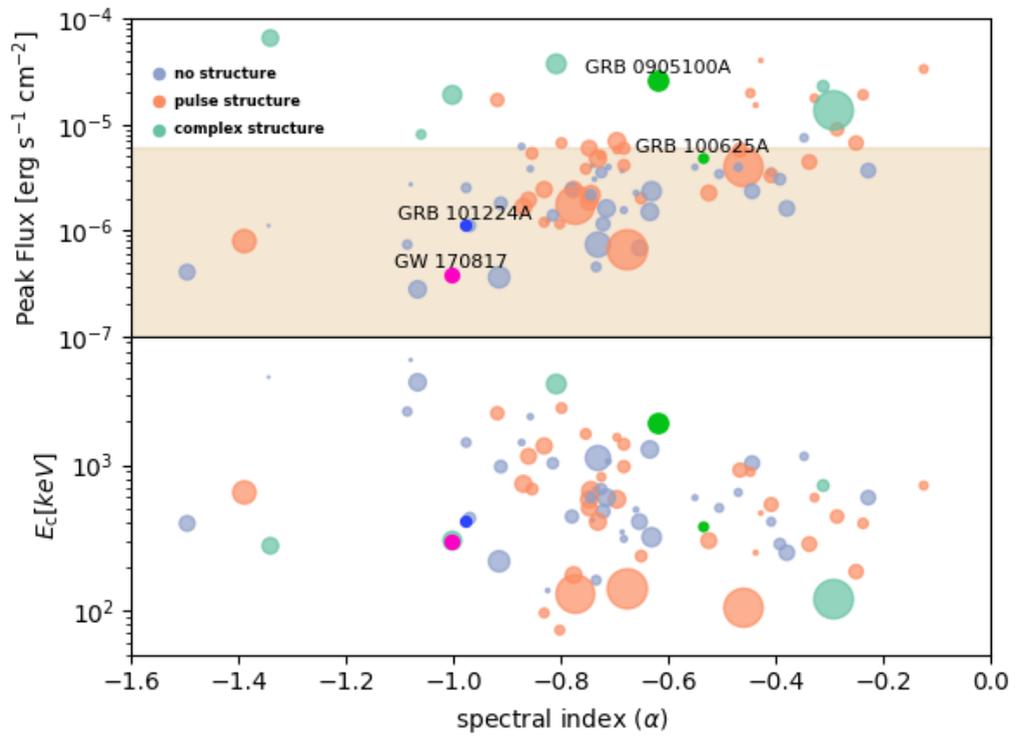
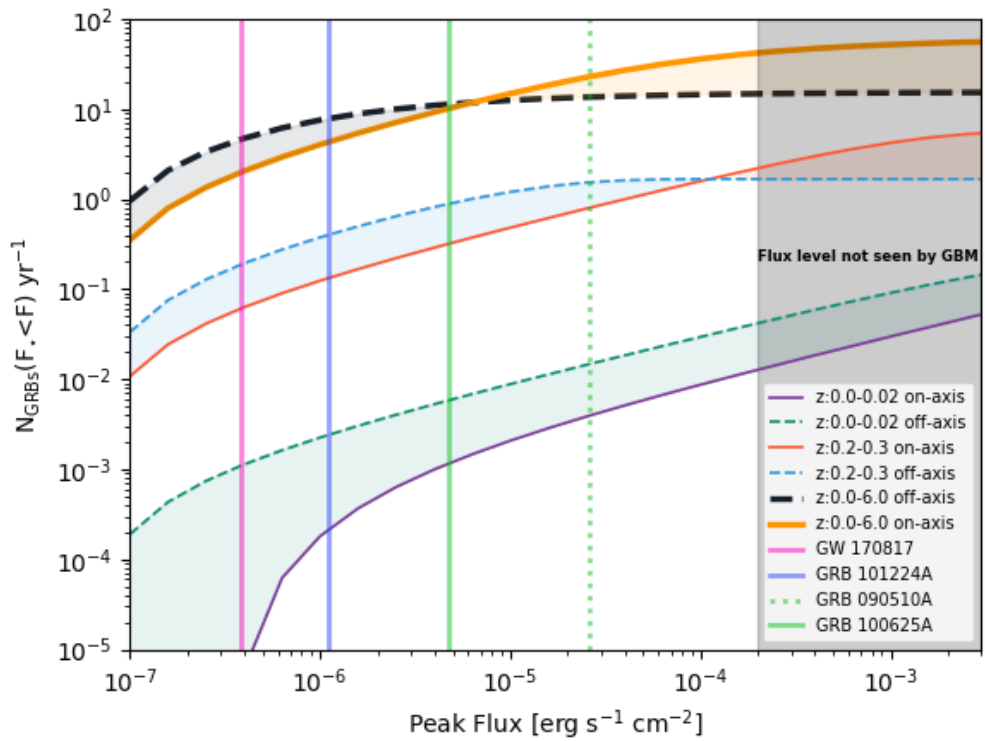

# Methods

## 1. Observations and Data Analysis

To obtain the spectral properties of the GRB associated to GW 170817, the temporal off-source photon event light curve was fitted with a polynomial via an unbinned Poisson likelihood. This polynomial was used to estimate the background counts during the source temporal interval. The on-source interval was derived by choosing bins via the Bayesian blocks method[28] and selecting bins with a 3σ excess over the estimated background. This resulted in a single bin lasting 0.65 s. The count spectra of this bin were fit via a Poisson-Gaussian likelihood to an exponentially cut-off power law resulting in a photon index of -1.0±0.32 and a cutoff energy of 230±130 keV. Spectral and temporal analyses were performed with the Multi-Mission Maximum Likelihood framework (3ML)[29].

All Fermi/GBM GRBs with a reported duration less than two seconds were reanalyzed with the above-mentioned method[30] and temporal bins were selected for spectral analysis if they exceeded a threshold of 3σ above background. To study the peak flux spectral properties of the population, we discarded events with errors greater than 50% on the photon spectral index and 60% on the cutoff energy to eliminate poorly fit spectra while simultaneously retaining those with constraints as good as those for GW 170817. The durations of the events were calculated by taking the start of the first significant bin to the end of the last significant bin which could include bins in between that were not significant.

GBM has observed 355 sGRBs to date[9]. After removal of poor fits and sGRB with badly modeled background in the GBM catalogs, we were left with 248 GRBs. Our cut on parameter constraints reduced the sample to 68 sGRBs. The choice of a cutoff power law for the spectral function does introduce a bias for low flux GRBs that would only be fit by a power law. These would not have proper flux estimates due to a lack of a measurable energy cutoff and are the main contribution to sGRBs that were removed from the sample due to poor constraints on the spectral parameters. Since these GRBs have low flux, our rate estimates are conservative.

Pulse structure was determined by dividing the observed significant intervals into three categories: (1) single bin, (2) contiguous bins with monotonic increase/decrease or rise and decay pulse shape and, (3) complex which including non-contiguous bins.

All data used is publicly available via the Fermi Science Support Center (FSSC) (https://fermi.gsfc.nasa.gov/ssc/data/access/gbm/).

## 2. Modeling of sGRB rates

We base our simple estimates of the rate on the method proposed previously[31,32]. We assume that the peak energy flux limit for detection by the GBM detector is $10^{-6}$ erg s$^{-1}$ cm$^{-2}$. Knowing the spectrum and the redshift, the maximum distance $d_{max}$ at which a given GRB is visible can be estimated (after k-correction). This maximum distance corresponds to a maximum volume. Then the rate is estimated as

$$R = \frac{1}{V_{max}} \frac{1}{\Omega} \frac{1}{T}$$

where Ω=0.5 is the fraction of the sky seen by GBM and T = 9 yr is the duration of the mission.

## 3. Estimate of ratio of on-axis to off-axis bursts

We assume that the angular-dependent, emitted, co-moving luminosity of a burst can be expressed as a step function

$$L_{em}(\theta_0, \theta_{obs}, L_0) = L_0 \quad \text{if } \theta_{obs} < \theta_0$$
$$L_{em}(\theta_0, \theta_{obs}, L_0) = L_0/F \quad \text{otherwise}$$

where $L_0$ is the luminosity emitted in the direction of the jet, $\theta_0$ is the opening angle, $\theta_{obs}$ is the angle between the jet's direction and the observer direction, and F is a reduction factor (taken to be 100). We further assume that if $\theta_{obs}$ is larger than 45°, the sGRB is not seen.

The distribution of emitted luminosity can be obtained by

$$\Phi(L_{obs}) = 2\pi \int d\theta_0 \xi(\theta_0) \int \sin(\theta_{obs}) d\theta_{obs} \int dL_0 \psi(L_0) \delta(L_{obs} - L_{em})$$

where $\psi(L_0)$ and $\xi(\theta_0)$ are the distributions of the jet's luminosity, $L_0$, and of the opening angle. We can take those distributions in the form

$$\psi(L_0) = A \left(\frac{L_0}{L_b}\right)^{-\beta} \exp\left(\frac{-L_b}{L_0}\right) H(L_{\text{lim}})$$

$$\xi(\theta_0) = B \exp\left(\frac{-\left(\log(\theta_0) - \log(\theta_{ref})\right)^2}{\sigma^2}\right)$$

where H is the Heaviside function, $L_b$ is the break luminosity which can be taken at few $10^{53}$ erg s$^{-1}$, $\beta = 0.5$ describe the decay of the luminosity distribution[33] and $L_{\text{lim}} = 5 \times 10^{50}$ erg s$^{-1}$ is the minimum luminosity of an on-axis burst. The opening angle distribution function is taken to be a log normal distribution centered on $\theta_{ref} = 7°$ with width $\sigma = 0.1$. In addition, A and B are normalization constants.

We further assume that the rate of sGRBs follows the retarded star formation rate[34]

$$R_{SGRB}(z) \propto \int C S_{FR} f(\tau(z, z')) \frac{dz}{dt'} dz'$$

where C is a normalization constant, SFR is the star formation rate at a given redshift and f is the distribution of delay τ, taken to be a log-normal distribution[35] centered on 3 Gyr with a deviation of 0.2 Gyr. The star formation rate is taken to be the fitting formula of the form

$$S_{FR}(z) = \frac{0.01 + 0.12z}{1 + \left(\frac{z}{3.23}\right)^{4.66}}$$

The rate of short GRBs between two given fluxes $F_1$ and $F_2$ up to a given redshift z can therefore be estimated as

$$N(z, F_1 < F < F_2) = R \int dz' \frac{dV}{dz'} R_{SGRB}(z') \int dF \Phi(4\pi d_L(z')^2)$$

where R is the local rate of short GRB, dV/dz is the comoving volume and dL is the luminosity distance. Overall, the rate is a 5-fold integral computed numerically with parallel processing on a GPU using CUDA.

## 4. Software availability

The software used is publicly available at https://github.com/giacomov/3ML.

## 5. Supplementary References

**Extended Data**

**Extended Data Figure 1 | Comparison of the gamma-ray spectrum of GW 170817 and GRB 101224A.**
The observer-frame νFν spectra of GW 170817 and GRB 101224 in comparison to the distant and likely on-axis GRB 090510A. The shaded regions indicate the 1σ confidence interval for the flux.

**Extended Data Figure 2 | Luminosity distribution.**
The isotropic luminosity of the peak flux bin for each GRB with redshift in our sample. Both local GRBs exhibit isotropic luminosities of about 1000 times less and spectra markedly softer those at larger distances. If we assume that high observed flux sGRBs are the result of on-axis emission, then we expect them to have harder observed spectra and high intrinsic luminosities.

**Extended Data Figure 3 | GBM peak flux distribution.**
To demonstrate the choices assumed for the minimum and maximum peak energy fluxes, we plot the peak energy flux of all and short GRBs from the official GBM catalog (REF). The shaded region is equivalent to the shaded region in Figure 2.

**Extended Data Figure 4 | Count rate light curve of GBM-GW170817.**
The Bayesian block binned count rate light curve of the GBM detected sGRB associated with GW 170817. The red line indicates the fitted background level and the shaded region indicates the times over which Bayesian blocks were applied. Additionally, the significance of the each temporal bin is displayed.

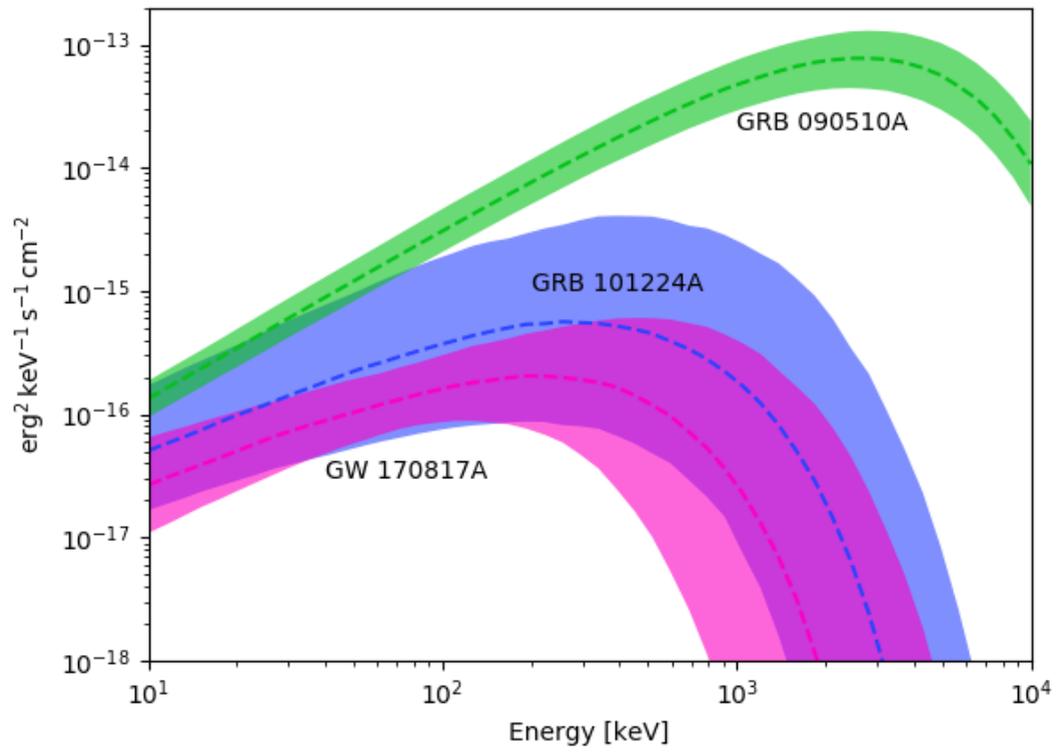

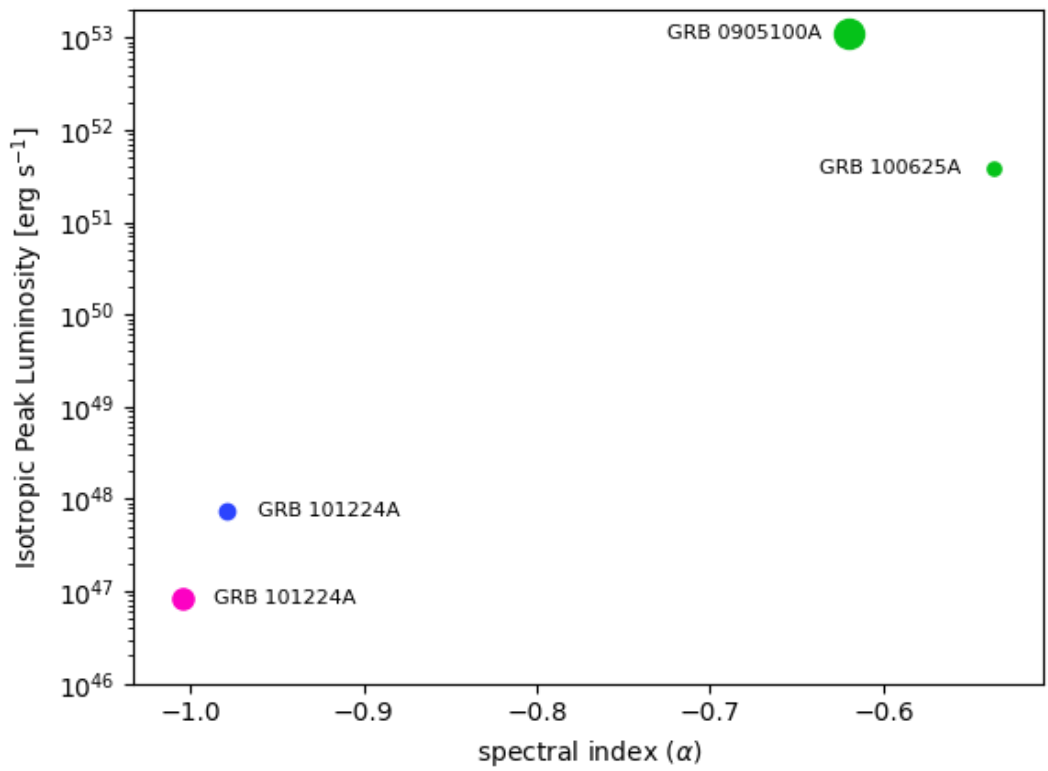

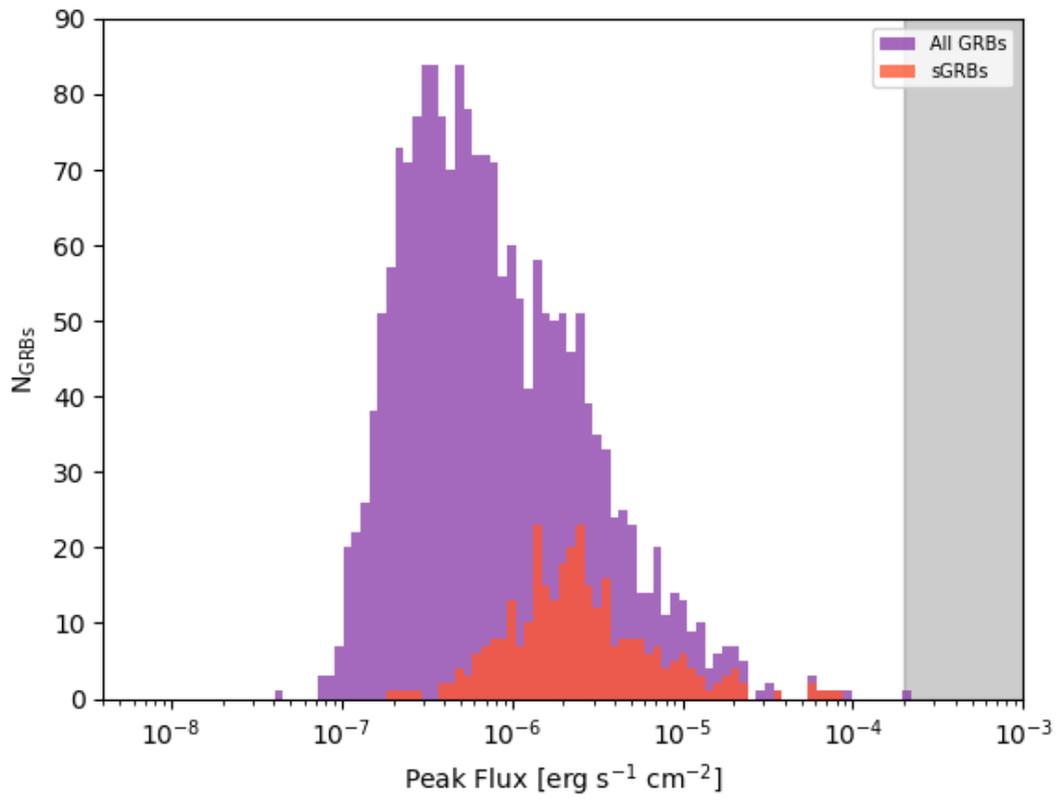

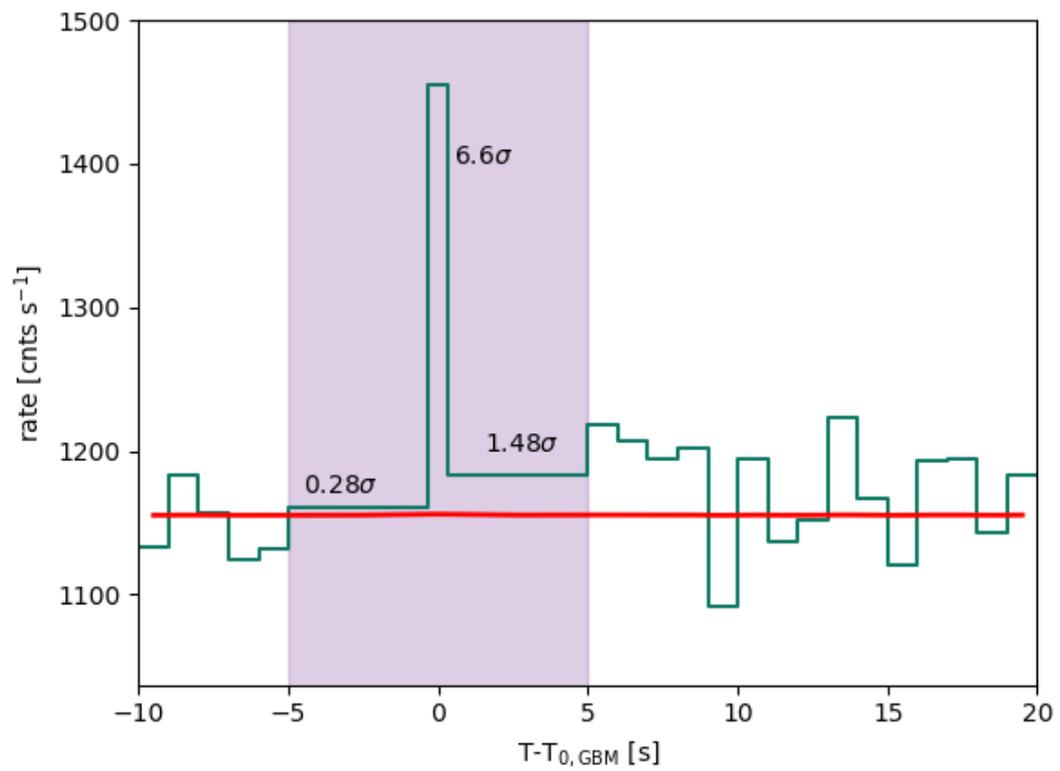